# *Caveats* for the Use of Citation Indicators in Research and Journal Evaluations

Loet Leydesdorff

Amsterdam School of Communications Research (ASCoR), University of Amsterdam,

Kloveniersburgwal 48, 1012 CX  Amsterdam, The Netherlands

<loet@leydesdorff.net> ; <http://www.leydesdorff.net>

**Abstract**

Ageing of publications, percentage of self-citations, and impact vary from journal to journal within fields of science. The assumption that citation and publication practices are homogenous within specialties and fields of science is invalid. Furthermore, the delineation of fields and among specialties is fuzzy. Institutional units of analysis and persons may move between fields or span different specialties. The match between the citation index and institutional profiles varies among institutional units and nations. The respective matches may heavily affect the representation of the units. Non-ISI journals are increasingly cornered into "transdisciplinary" Mode-2 functions with the exception of specialist journals publishing in languages other than English. An "externally cited impact factor" can be calculated for these journals. The citation impact of non-ISI journals will be demonstrated using *Science and Public Policy* as the example.

# 1. Introduction

The emergence of new search engines for citation analysis like *Google Scholar* and Elseviers's *Scopus* (Meho & Yang, forthcoming) and the increased emphasis on quality control using the list of journals included in the citation indices of the Institute of Scientific Information (Thomson-ISI) has led to the emergence of an evaluation industry. The US National Science Foundation recently launched a new program for the study of science and innovation policies (at http://www.nsf.gov/pubs/2007/nsf07547/nsf07547.htm). Science indicators also figure prominently in the Seventh Research Framework Program (FP7) of the European Commission. Furthermore, evaluations using publication and citation counts have become standard practices in national research assessment exercises and evaluations at university, faculty, and even departmental levels.

This increased interest in bibliometric indicators puts an onus on scientometricians. Since earning-power is increasingly important for university positions and tenure may depend on new contracts, relations with the clientele have become business-like and strategic in this field (Gibbons *et al*., 1994; Nowotny *et al*. 2001). Entry barriers are based on existing relations of trust. Newcomers on the supply side profile themselves with new and additional ranking and search options. However, the results are sometimes conflicting and confusing because of different parameter choices (Anderson *et al*., 1988; Leydesdorff & Wagner, 2007).

For example, the *Journal-ranking.com* at the Internet launched a new Center for Journal Ranking (CJR) in April 2006. This tool builds on the underlying information provided by the *Journal Citation Reports* (JCR) of the Institute of Scientific Information (ISI). While the traditional impact factors (Sher & Garfield, 1965; Garfield, 1972) use a two-year time window,[1] the CJR enables the user to search impact factors using different periods of time (Garfield, 1998a; Frandsen & Rousseau, 2005) and to rank journals within subject categories. Both the time-frames of the citation measures and the differences in citation practices among fields of science are major sources of variation. Thus, a parameter space is generated in which an analyst has to make reasoned and/or pragmatic choices.

When developing the impact factor, Garfield (1972) originally opted for a shorter time-frame because he thought that a two-year time limit provided a sufficient sample on the basis of work of Martyn & Gilchrist (1968). Garfield (2003) admitted that another reason for the two-year period was that it suited biochemistry and molecular biology—major areas of interest for the ISI. The impact factor normalizes for size because of the division by the number of publications in the denominator and thus gives small journals a better chance of selection. Bensman & Wilder (1998) provided evidence that total citations (i.e., accumulated impact) correlate with perceived quality of journals more than the impact factors (Leydesdorff, 2007a). Total citations can be considered as reflecting the prestige of a journal, while impact factors highlight a journal's current value at one or more research fronts. The two measures are correlated (Spearman's $\rho = 0.73$; $p < 0.01$; JCR 2005).² Total citations, of course, are more stable over time—since accumulating—than impact factors.

---

[1] The impact factor for a journal is calculated by dividing the number of current year citations to the source items published in that journal during the previous two years. It can be considered to be the average number of times published papers are cited up to two years after publication (Garfield, 1979, at p. 149).

Recently, Hirsch (2005) has proposed the *h*-index as a new indicator for evaluation purposes. This index is defined as the highest rank on a scientist's list of publications such that the first *h* publications received at least *h* citations. (The indicator follows naturally from the ordering of the listings using Google Scholar because the search results are approximately ranked in terms of the number of citations. In order to determine the *h*-index, one scrolls down to the sequence number which equals the number of citations of the author.) The *h*-index can be extended to any set of documents; for example, journals (Braun *et al.*, 2005). In the meantime, the *h*-index has also been included under the so-called "Citation Reports" available online at the ISI's Web-of-Science. For example, 162 documents can be retrieved as published in *JASIST* in 2005. The *h*-index of this set is six, while the index is ten for the 139 documents published by *JASIST* in 2004.³ Unlike the impact factor, the *h*-index cannot decrease for a given set and can thus be considered as an accumulating indicator for lifetime achievement in the case of individual scholars.

As in the case of the impact factor (Egghe & Rousseau, 1990; Moed, 2005; Rousseau, 2004; cf. Seglen, 1997), scientometricians have been prolific in proposing a series of technical improvements to the measurement of the new indicator. Egghe (2006a, 2006b) first developed the *g*-index, Jin (2006) followed with the *A*-index, and, more recently, Jin

---

² When the *Social Science Citation Index* is excluded from this set, the Spearman's $\rho = 0.71$; $p < 0.01$.
³ These measurements were done on July 27, 2007.

*et al*. (2007) suggested correcting the *h*-index for the aging of *papers* using the *AR*-index. Burrell (2007) proposed the *h*-rate which intends to normalize the *h*-index of an *author* for one's career length. For pragmatic reasons, Kosmulski (2006) proposed to use not the *h*, but $h^2$ for cutting the list of (*h*) cited papers in fields of science (like medicine and biology) where the number of citations per article is high.

As in the previous refinements of the impact factor, in my opinion, mathematical sophistication cannot increase the validity of an indicator used in processes of research evaluation. Most importantly, exogenous variation is caused by different citation and publication practices among fields of science (Leydesdorff & Amsterdamska, 1990), but also within fields of science among different publication channels such as reviews, research articles, and letters (Garfield, 1996, 1998b; Bensman, 2007). For example, Garfield (1980, at p. 1A) summarized this conclusion as follows: "I've often stressed the importance of limiting comparisons between journals to those in the same field." Most evaluators pay lip-service to this problem by stating it up front, but then propose a pragmatic solution and proceed to the measurement. Some major research teams have developed their own (in-house) classification schemes which they claim to be robust (Pinski & Narin, 1976; Moed *et al.*, 1985; Glänzel & Schubert, 2003).

For lack of an agreed-upon alternative, the ISI subject categories are often used for "comparing like with like" (Martin & Irvine, 1983). These categories are assigned by the ISI staff on the basis of a number of criteria, including the journal's title, its citation patterns, etc. (McVeigh, *personal communication*, 9 March 2006). The classifications, however, match poorly with classifications derived from the database itself on the basis of analysis of the principal components of the networks generated by citations (Leydesdorff, 2006a, at pp. 611f.). Using a different methodology, Boyack, Klavans & Börner (2005) found that in somewhat more than 50% of the cases the ISI categories corresponded closely with the clusters based on inter-journal citation relations. These results accord with the expectation: many journals can be assigned unambiguous affiliations in one core set or another, but the remainder which is also a large group is very heterogeneous (Bradford, 1934; Garfield, 1972).

In summary, I listed a number of (partly well-known) problems which I will address below using empirical data. In the next section (section 2), I show that impact factors are different with an order of magnitude when comparing journals in mathematics with journals in genetics. In section 3, I address the other major source of variation in the database: the different time-frames of publication media. Journals which publish letters are compared with those that publish exclusively reviews. The two types of media exhibit significantly different cited journal half-life patterns. Therefore, even within a narrowly defined specialty, one cannot expect publications in these different media to have comparable long- and short-term citation impacts.

Elsewhere (Leydesdorff, 2006a), I have shown that journals cannot unambiguously be classified into journal categories on the basis of their aggregated citation patterns. All sets are overlapping and fuzzy (Bradford, 1934; Garfield, 1971; Bensman, 2007). When one evaluates research programs and research groups on the basis of articles, one always needs a reference set (Studer & Chubin, 1980, at p. 269). However, decisions about how to cut the cake determine what is included and what is excluded. Furthermore, institutional research programs and even creative individuals can be expected to cross disciplinary boundaries (section 4; cf. Kreft & De Leeuw, 1988).

The preliminary delineation remains that of the selection of journals included in the ISI-databases versus the "non-ISI journals" (Garfield, 1972). This is the subject of section 5. What are the effects of the increasing pressures at the institutional levels to publish exclusively in "ISI journals"? Using *Science & Public Policy* as an example, I shall show what the inclusion of this journal adds to the network of journal-journal citation relations. The non-ISI journals can be retrieved within the ISI-set in terms of their being-cited patterns. What are the effects of exclusion of such a journal on the relevant definitions of specialties?

**2. Variation in the impact factors among fields of science**

Figures 1 and 2 show the distribution of impact factors for two subject categories of the ISI: "mathematics" ($N = 191$ journals) and "genetics" ($N = 133$). As noted, the journal selection and the ISI-categories are themselves debatable, but for the purpose of this argument the comparison between these two sets is revealing. The values of the *x*-axis differ not slightly, but by an order of magnitude!

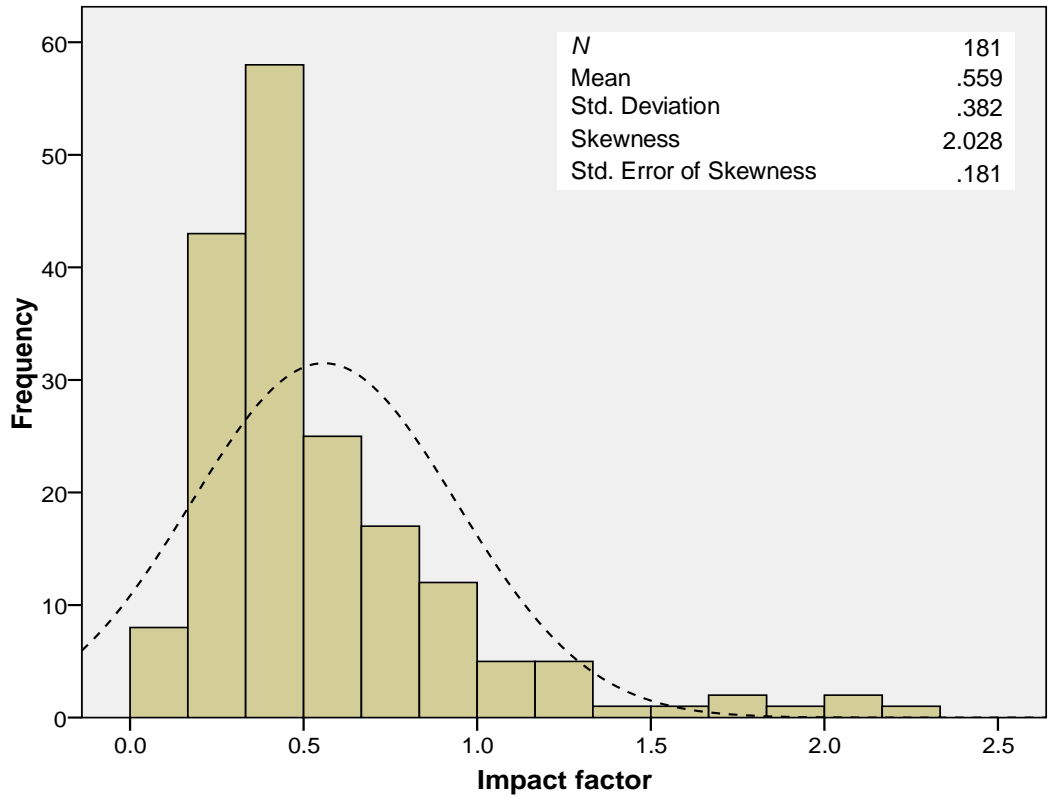

**Figure 1**: Impact factors of 181 journals in Mathematics according to the classification of the ISI (source: *Journal Citation Reports* 2005).

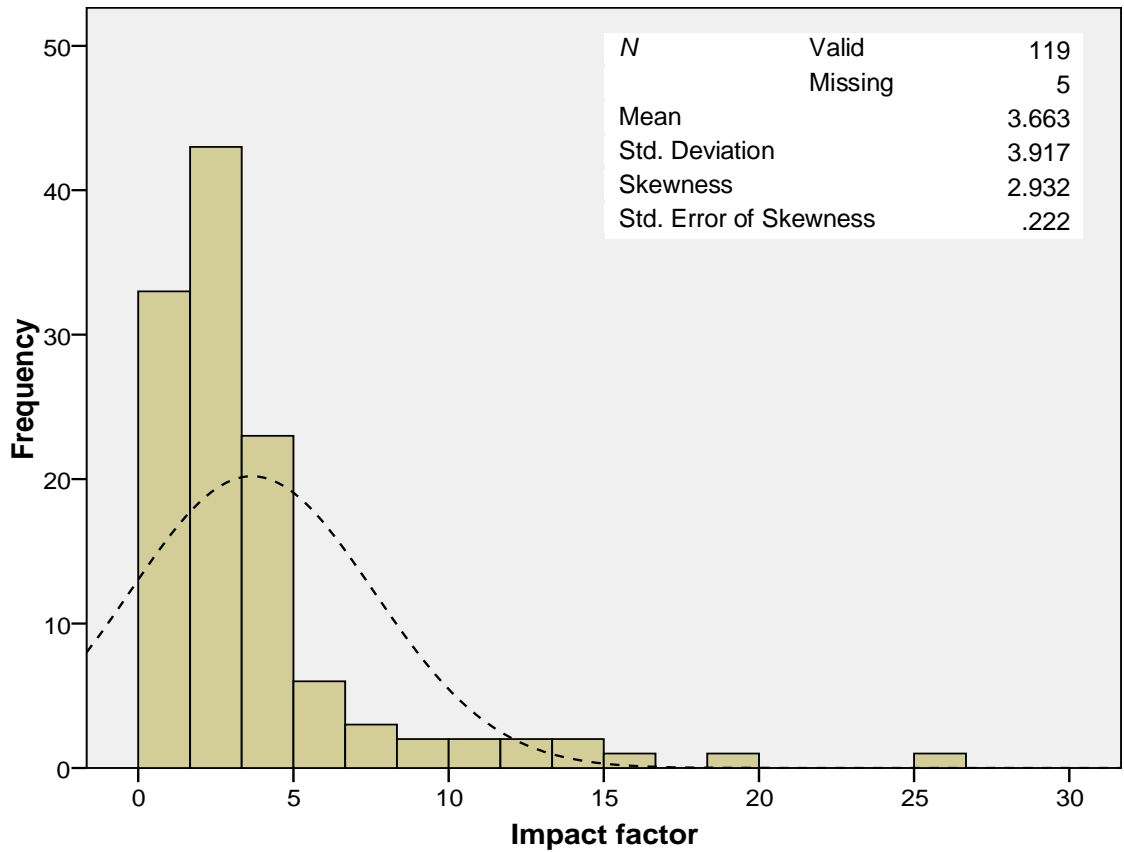

**Figure 2**: Impact factors of 119 journals in Genetics according to the classification of the ISI (source: *Journal Citation Reports* 2005).

The problem is well-known among scientometricians (e.g., Martin & Irvine, 1983). However, it has an important policy implication: one is not allowed to compare evaluations based on citation rates across disciplines. Consequently, it is not legitimate to allocate funds across fields of science on the basis of comparisons among citation rates for different disciplines (National Research Council, 2007). In other words, one has to disaggregate. The rules for the disaggregation, however, are far from obvious.

## 3. Variation in the cited journal half-life among journal types

In addition to the static variation in publication and citation behavior among fields of science, the impact factor and the recently proposed modifications of the *h*-index (Jin *et al.*, 2007; Burrell, 2007) incorporate the time dimension. All indicators imply a choice for a time-frame, but sometimes this choice is left implicit. Scientometricians often use a so-called citation window for the standardization (Moed, 2005). As a standard for the aging effect of journal publications, the ISI provides the so-called "cited half-life" of a journal. The cited half-life for the journal is the median age of its articles cited in the current year. In other words, half of the citations to the journal are to articles published within the cited half-life (Rousseau, 2006b). One can expect these measures also to differ among fields of science (Price, 1970). In the humanities and the social sciences, research fronts are often virtually absent (Nederhof *et al.*, 1989).

Independently of the differences among fields of science, publications come in different types: articles, letters, and reviews. This source of variation can be expected to influence the measurement of citation rates across fields of science. By using a relatively short time window, one can expect to favor letters as compared with other publications. Similarly, the ranking of reviews and review journals is negatively affected by the decision to focus on the dynamics of the last two years at research fronts. Accumulating indicators like measures for life-long achievements tend to work the other way round. The various types of publications have both different impacts and life-cycles.

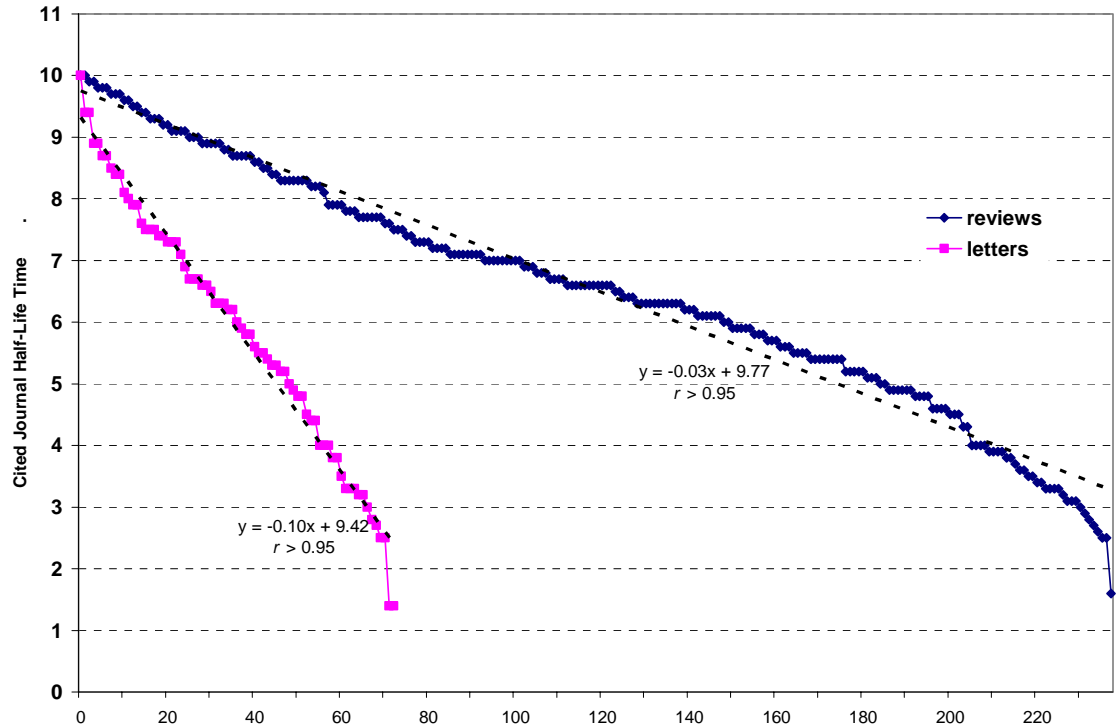

**Figure 3**: Descending Cited Journal Half-Life of 238 journals with "Review" and 73 with "Letters" among their title words (source: *Journal Citation Reports*, 2005).[4]

Figure 3 substantiates this point empirically by showing the effects of sorting letters and review journals in terms of their cited half-life in descending order (using the *Journal Citation Reports* of 2005). A one-way analysis of variance using the two subsets of journals with the words "Reviews" or "Letters" in their titles, versus the third subset of remaining journals shows that the impact factors of review journals is significantly higher ($p < 0.01$) than those of other journals (Table 1). The cited half-life times of review journals are not significantly higher than those of other journals at the one-percent level, but the cited half-life times of letters are on average more than twice as low. As can be expected, the correlation between impact factors and cited half-life times over the file is negative: $\rho = -0.239$ ($p < 0.01$; JCR2005).

---

[4] Additionally, 90 journals with "Review" in their title and three with "Letters" have no cited half-life times listed in the *JCR* 2005 because they extend beyond 10 years. Four more journals with "Letters" in their title have no cited half-times listed.

|  |  | N | Mean | Standard Deviation | Standard Error | 95% Confidence Interval for Mean | |
|---|---|---|---|---|---|---|---|
|  |  |  |  |  |  | Lower Bound | Upper Bound |
| Impact Factor | "Reviews" | 326 | 4.0081 | 6.324776 | 0.350297 | 3.31897 | 4.69724 |
|  | "Letters" | 79 | 1.79813 | 1.902339 | 0.21403 | 1.37203 | 2.22423 |
|  | Other journals | 7386 | 1.47353 | 2.022984 | 0.023539 | 1.42739 | 1.51967 |
|  | Total | 7791 | 1.58287 | 2.41719 | 0.027385 | 1.52919 | 1.63656 |
| Cited Half-life | "Reviews" | 294 | 25.24932 | 37.507056 | 2.187456 | 20.9442 | 29.55444 |
|  | "Letters" | 76 | 9.57632 | 18.5418 | 2.12689 | 5.33933 | 13.8133 |
|  | Other journals | 6790 | 24.22607 | 36.954594 | 0.44847 | 23.34693 | 25.10521 |
|  | Total | 7160 | 24.11258 | 36.858479 | 0.435593 | 23.25869 | 24.96647 |

**Table 1**: One-way Analysis of Variance of the impact factors and cited half-life times for journals with the words "Reviews" or "Letters" in their titles, compared with the remainder of the JCR-set.

In summary, at the level of the ISI-set, the major problems in evaluation are the delineation and the assumption of homogeneity in the sets to be evaluated. All sets are overlapping and fuzzy (Bradford, 1934; Garfield, 1971; Bensman, 2007). Within relatively homogenous sets, different media of communication (articles, letters, and reviews) can be expected to generate different citation dynamics. For analytical purposes, one may wish to confine studies in the sociology of science to homogenous sets like only research articles in a restricted set of core journals, but for the practical purpose of evaluation such analytical restrictions may be counterproductive.

**4. Institutions and individuals**

Unlike analytical units of analysis like disciplines and specialties, social institutions have the function to integrate different perspectives in research programs. If one wishes to compare—or even benchmark—institutions in terms of citation rates, the above figures suggest that a university could be advised to close down its mathematics department in order to increase its standing, for example, on the national ranking (Leydesdorff & Zhou, 2005). In summary, institutions are the wrong units of analysis for scientometric comparisons in terms of delineated journal sets because they are intellectually heterogeneous (Collins, 1985).

Recently, Braam (2007) proposed to use "journal scope change" in the publication profile of research programs and/or research groups as an indicator for intellectual growth. However, this begs the question of how to calibrate change at the level of a group against change at the level of relevant fields (Whitley, 1984; Bonaccorsi, 2005). If a group rows in terms of its publications against the tide at the journal level, chances of success are different from when one has an opportunity to ride the wave of new developments (Leydesdorff & Van der Schaar, 1987; Rafols & Meyer, 2007). In an established specialty, it may be a more rewarding strategy to focus on core journals or, in other words, to be less innovative in terms of exploring interdisciplinary relations. Furthermore, the trade-off may be different for individuals, research groups and/or (national) research programs.

Nations can be considered as aggregated institutional units of analysis. Thus, what holds for institutions holds also for comparisons among nations (King, 2004). For example, the research portfolios of East Asian countries are less developed in the bio-sciences than those of the USA and most Western European countries (Park *et al*., 2005). The ISI database, however, is more strongly developed in the life sciences than in the physical sciences. Thus, the match of East Asian countries with the database is weaker than for some Western countries. The latter will for this reason enjoy an advantage in terms of their visibility (and thus, their chances of being cited) within the database.

While nations are macro-units of institutional analysis, individuals can perhaps be considered as minimal units of analysis. Individuals may also be heterogeneous in terms of their scientific output (Hellsten *et al*., forthcoming). In a certain sense, creative researchers are supposed to generate new variations or, in other words, knowledge claims that reach beyond existing borderlines. Bar-Ilan (2006) found that Google Scholar reflected this heterogeneity more than the ISI-databases in the case of the information sciences. However, one is not allowed to apply statistics as a predictor in individual cases. Individuals almost by definition deviate from the mean.

Finally, a theoretical reflection is here in place. The prevailing conceptualization in the sociology of science in terms of levels (group versus field; context of discovery versus context of justification) misses the communication dynamics between these two dimensions of the scientific enterprise. Publications contain knowledge claims that compete for proving their value at the level of (one or more) scientific discourses. Citations can be considered indicators of diffusion at the network level and cannot inform us about the intrinsic quality of research at the site of production. Knowledge claims in publications provide the variation.

Publication counts can inform us about the *quantity* of the production. Citation counts reflect the usefulness of the publication in the construction of new knowledge claims at *other* places in the network (Fujigaki, 1998). However, networks of citations among journals are structural (Price, 1965). They are reproduced from year to year and can thus be expected to exhibit their own dynamics. It is not only the intrinsic quality of specific publications, but also their position and timing in the distribution that can be expected to determine how the networks will absorb new variants (Leydesdorff, 1998).

**5. Non-ISI journals**

Many editors and publishers feel a sense of frustration if their journal is not included in the ISI dataset. The ISI provides a number of criteria for inclusion such as (1) three consecutive issues must be published on time; (2) how many citations the journal has had in the previous two years in other journals of the *Science Citation Index* or the *Social Science Citation Index*, respectively; and (3) special factors, such as journals which appeal especially to decision-makers. The latter users are considered influential, but they do not write academic papers with systematic references (Page, *personal communication*, 16 February 2007). The criteria are weighted and this procedure is a company secret of the ISI (Garfield, 1990; Testa, 1997). As can be expected, the ISI is under pressure from lobbying agencies (Moravcsik, 1985; Arvinitis & Gaillard, 1992) and publishing houses to include journals on various grounds (Tijssen & Van Leeuwen, 1995; Hicks, 1999; Maricic, 1997).

Inclusion in this database has become increasingly important for journals in recent years because of the strong pressures on institutions and authors, particularly in Asia, to publish in journals with impact factors that are included in the ISI-databases. Using the ISI-database, however, it is possible to calculate a quasi impact factor for journals not included in the ISI set because all the citations—including those to non-source journals—are part of the database (Stegmann, 1997; 1999). This so-called "externally-cited impact factor" (Christensen *et al.*, 1997, at p. 536) underestimates the real impact factor because in this case one does not include "within-journal self-citations" (Leydesdorff, 2007a).[5] This can lead to a considerable under-representation: 124 of the 299 citations (41.5%) in 2005 to the 2003 and 2004 volumes of *Scientometrics*, for example, were provided by authors publishing in this same journal. For *JASIST*, this percentage was 20.8% in 2005.

In order to construct this externally-cited impact factor, one can search on the Web of Science, for example, with "Sci Publ Policy"—that is, the abbreviation for *Science and Public Policy*—as the cited work, a specific year (e.g., 2005) for the citations, and the two preceding years for the cited years. This provides not the citation score, but the number of unique article linkages generated by citations. However, each citing article may cite different sources from the same journal. By aggregating all the citations from the citing journals to the cited journal using the JCR, one can then construct an impact factor. This routine can also be automated. The externally-cited impact factor 2005 for *Science and Public Policy* is (25/79 =) 0.316.[6] (Let me follow the standard practice of providing impact factors with three decimals although one should be aware that this impressive precision is part of the codification process surrounding impact factors.)

---

[5] Within-journal self-citations are not necessarily self-citations by authors, but they provide us with a measure of how much a journal is used by a scientific community for an inward-directed discourse (Leydesdorff, 2007a and b).
[6] The number of citable issues was kindly provided by the publisher of *SPP*.

The citation impact environment of *Science & Public Policy* is visualized in Figure 4 using techniques developed in other contexts (Leydesdorff, 2007a and b). *SPP* is cited in 2005 by articles in 29 other journals. Twenty journals are included in its citation environment above the threshold of cosine = 0.2.[7] The figure is interesting because it provides a more integrated view of the field of Science & Technology Studies (STS) than one is able to generate by using one of the leading journals of the field as a seed journal.

---

[7] The cosine is increasingly used as a similarity criteria because unlike the Pearson correlation the normalization is not to the arithmetic, but to the geometrical mean. On the one hand, this solves the problem of highly skewed (i.e., non-normal) distributions, and, on the other hand, the cosine values can be taken as input to the vector-space-model for the visualization (Salton & McGill, 1983; Ahlgren *et al.*, 2003; Leydesdorff, 2007a).

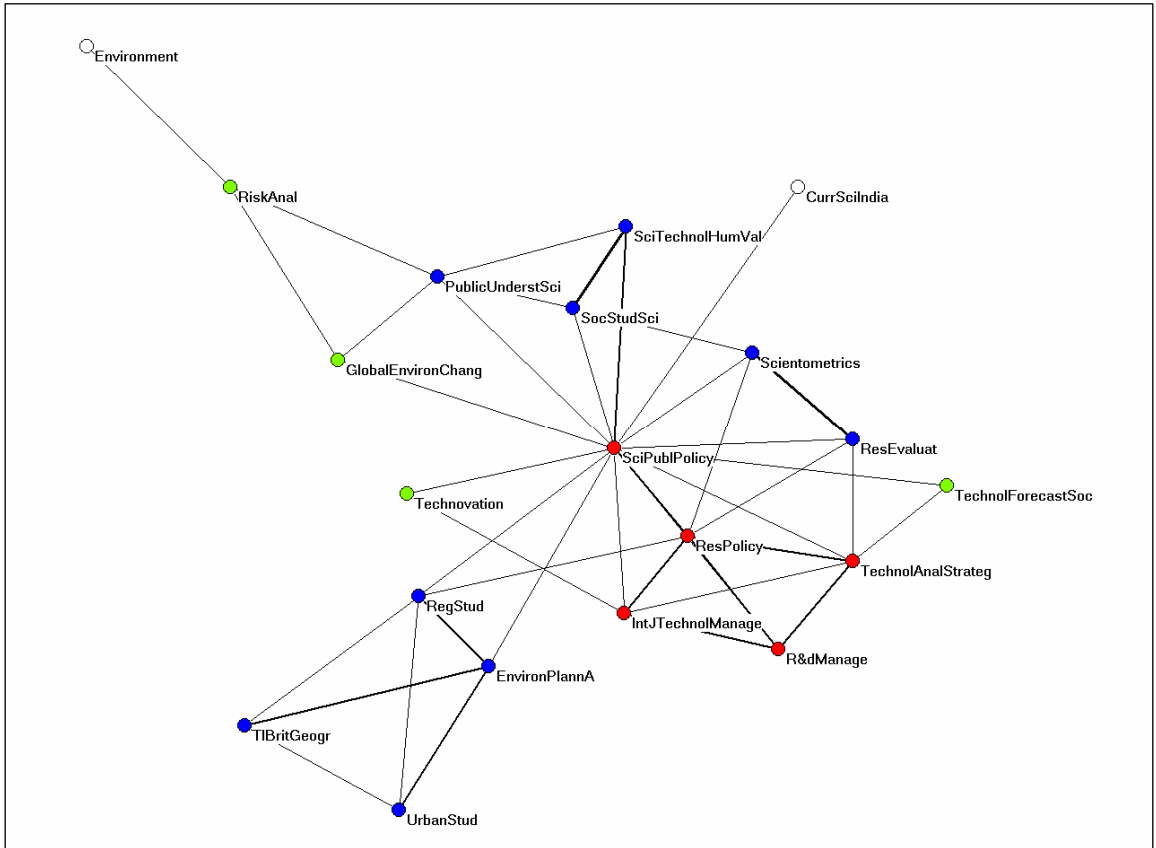

**Figure 4**: The citation impact environment of *Science and Public Policy.* (JCR 2005; *N* = 20; cosine ≥ 0.2).

During the last decade, the journal *Scientometrics* has become increasingly embedded in the information sciences (Figure 5). Unlike *JASIST* which is extensively cited in both information and computer science journals (to a total of 134 journals in 2005), *Scientometrics* has remained a specialist journal at the interface between information science and STS (cited only in 34 other journals in 2005). Two other core journals of STS—*Social Studies of Science* and *Research Policy*—have hardly any citation traffic between them. As increasingly an *interdiscipline*, STS is oriented towards and cited by journals in a variety of other disciplines (Leydesdorff & Van den Besselaar, 1997; Van den Besselaar, 2001; Leydesdorff, 2007b).

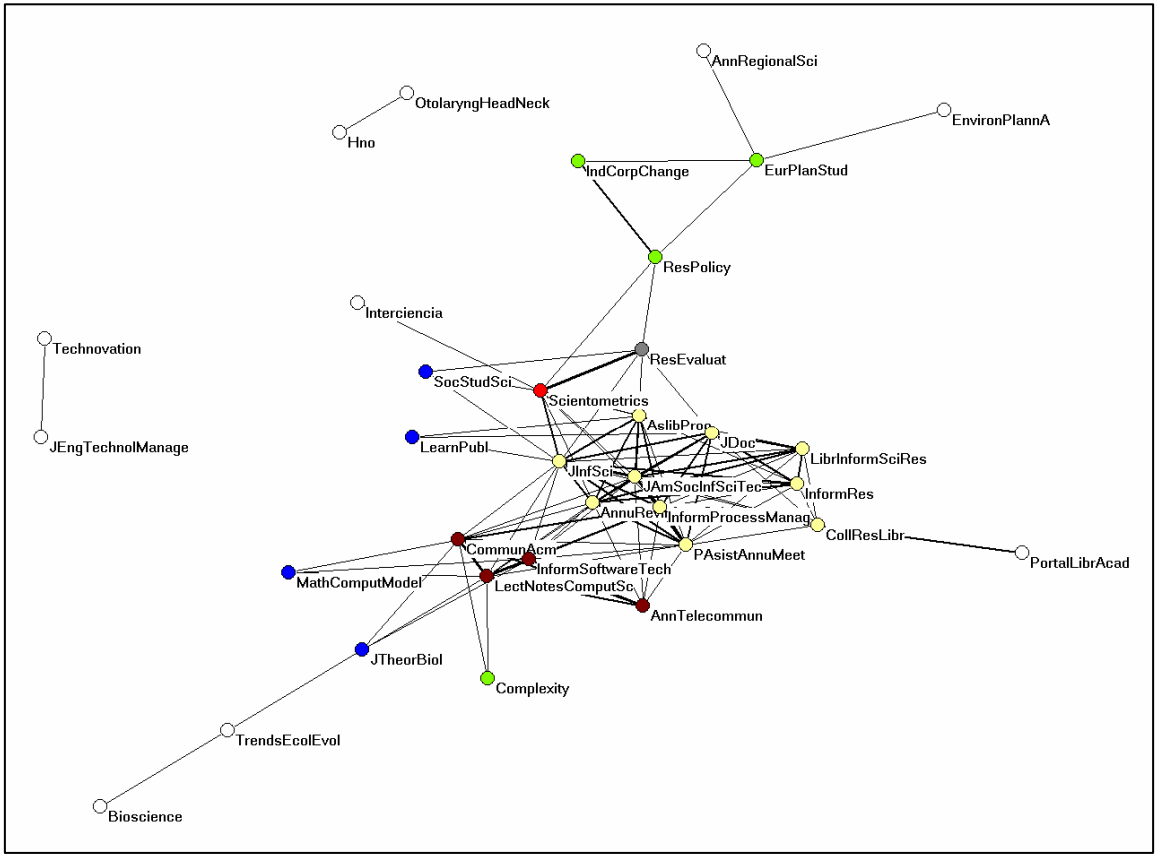

**Figure 5**: The citation impact environment of *Scientometrics*. (JCR 2005; $N$ = 34; cosine $\geq 0.2$).

Table 2 provides the values of the impact factors and the externally-cited impact factors—corrected for within-journal self-citations—for the 20 journals in the citation impact environment of *SPP*. The impact factor of *SPP* remains the lowest among them also after correction for "within-journal self-citations." This is a consequence of being a central node in an otherwise weakly connected network. However, the "betweenness centrality" of *SPP* in this environment is 69.9%, while this value is only 4.4% for *Scientometrics* in the ego-network of this journal as depicted in Figure 5. In other words, *SPP* fulfills an important function in this interdisciplinary network of journals, but this is not reflected in its impact factor because of the thinness of this network (Freeman, 1977; Hanneman & Riddle, 2005; Leydesdorff, 2007c).

| 2005 | Impact factor (ISI) | Externally-cited impact factor |
|---|---|---|
| *Current Science* | 0.728 | 0.492 |
| *Environment and Planning A* | 1.367 | 1.222 |
| *Environment* | 1.020 | 0.571 |
| *Global Environmental Change--Human and Policy Dimensions* | 1.952 | 1.629 |
| *International Journal of Technology Management* | 0.240 | 0.194 |
| *Public Understanding of Science* | 0.913 | 0.739 |
| *R & D Management* | 0.506 | 0.454 |
| *Regional Studies* | 1.525 | 1.258 |
| *Research Evaluation* | 0.474 | 0.421 |
| *Research Policy* | 1.835 | 1.125 |
| *Risk Analysis* | 1.510 | 1.217 |
| *Science & Public Policy* | n.a. | 0.316 |
| *Science Technology & Human Values* | 1.439 | 1.366 |
| *Scientometrics* | 1.738 | 1.017 |
| *Social Studies of Science* | 0.929 | 0.768 |
| *Transactions of The Institute of British Geographers* | 2.218 | 1.927 |
| *Technology Analysis & Strategic Management* | 0.446 | 0.410 |
| *Technological Forecasting and Social Change* | 0.811 | 0.444 |
| *Technovation* | 0.497 | 0.377 |
| *Urban Studies* | 0.988 | 0.752 |

**Table 2**: ISI-Impact factors and quasi impact factors of journals in the citation environment of *Science & Public Policy*.

It is not possible to generalize to all non-ISI journals because this is a very heterogeneous set, including among others also newspapers. The *Journal Citation Reports* 2005 contain 282,955 references to non-ISI sources. The analysis was hitherto only done on a case-by-case basis. On the basis of a large number of case studies of these non-ISI journals (Bornmann et al., 2007; Zhou & Leydesdorff, 2007a and b), I conjecture that scientific journals which are not included in the ISI database—that is, the so-called B-journals in evaluation studies—fall increasingly into two major groups. First, there are journals which like *SPP* fulfill transdisciplinary functions, but—unlike general journals such as *Science* and *Nature*—at the micro-level and on the margins of different specialties. Among the medical journals, these can, for example, be journals which focus on specific diseases. For example, the *Journal of Alzheimer's Disease* has recently been included in the *SCI-Expanded* at the web, but is not yet included in the JCR-set.

The second group consists of disciplinary journals which were selected away in favour of other journals in the same cluster. These may include journals publishing in languages other than English,[8] since virtually all papers in the *Science Citation Index* nowadays are in English (Table 3).

---

[8] Journals publishing in both English and other languages can be overrepresented in the database because authors sometimes cite both editions. Bornmann *et al.* (2007), for example, found an overestimation of impact factor of *Angewandte Chemie of* 21.5% because of this effect.

|  |  | Frequency | Percent | Valid Percent | Cumulative Percent |
|---|---|---|---|---|---|
| Valid | English | 998635 | 98.7 | 98.7 | 98.7 |
|  | German | 4941 | .5 | .5 | 99.2 |
|  | French | 2368 | .2 | .2 | 99.5 |
|  | Chinese | 2074 | .2 | .2 | 99.7 |
|  | Spanish | 1156 | .1 | .1 | 99.8 |
|  | Japanese | 926 | .1 | .1 | 99.9 |
|  | Russian | 861 | .1 | .1 | 100.0 |
|  | Czech | 165 | .0 | .0 | 100.0 |
|  | Multi-Lang | 83 | .0 | .0 | 100.0 |
|  | Finnish | 62 | .0 | .0 | 100.0 |
|  | Portuguese | 24 | .0 | .0 | 100.0 |
|  | Romanian | 18 | .0 | .0 | 100.0 |
|  | Latvian | 14 | .0 | .0 | 100.0 |
|  | Welsh | 12 | .0 | .0 | 100.0 |
|  | Italian | 10 | .0 | .0 | 100.0 |
|  | Slovak | 6 | .0 | .0 | 100.0 |
|  | Afrikaans | 3 | .0 | .0 | 100.0 |
|  | Dutch | 2 | .0 | .0 | 100.0 |
|  | Danish | 1 | .0 | .0 | 100.0 |
|  | Gaelic | 1 | .0 | .0 | 100.0 |
|  | Serbian | 1 | .0 | .0 | 100.0 |
|  | Total | 1011363 | 100.0 | 100.0 |  |

**Table 3**: language distribution of papers included in the CD-Rom version of the *Science Citation Index 2005*.

Perhaps, it may be increasingly difficult for journals that publish in English and are not included in the ISI database to maintain a position within a disciplinary setting. A division between disciplinary ("Mode-1") and transdisciplinary ("Mode-2") journals might thus further be reinforced (Gibbons *et al*., 1994; Leydesdorff & Jin, 2005).

## 6. Discussion and conclusions

Can journals and journal rankings be used for the evaluation of research? Garfield's (1972 and 1979) original purpose when creating the impact factor and the *Journal Citation Reports* was not to evaluate research, but journals! Based on Bradford's (1934) Law of Scattering, Garfield (1971) formulated his Law of Concentration which states that the tail of the literature of one discipline consists, in a large part, of the cores of the literature of other disciplines (Garfield, 1979, at p. 23): "So large is the overlap between disciplines, in fact, that the core literature for all scientific disciplines involves a group of no more than 1000 journals, and may involve as few as 500."

Thus, the ISI journal selection is based on its purposefulness for information retrieval. A journal which links literatures together, like the above example as *SPP*, may demonstrate the "strength of weak ties" (Granovetter, 1973), but the intellectual and social organization of the sciences is not the primary concern of the providers of this database. The impact factors provide a summary statistics which conveniently allows for the ranking of journals. However, these statistics are based on the means of a highly skewed distribution. Elsewhere, Leydesdorff & Bensman (2006) explained how the organization of the journal set in subsets can be expected to lead to a compounded distribution: each of the subsets of a scientific journal set has different underlying probabilities and therefore a different expected value or arithmetic mean. I showed this empirically for the cases of mathematics and genetics: the means differed with an order of magnitude. In the case of compounded distributions, heterogeneity and contagion act multiplicatively instead of additively, creating exponential and curvilinear relationships instead of the assumed additive, linear ones. In other words, an impact factor of two is not twice as good as an impact factor of one. Much depends on the context(s) of the measurement.

| Incorrect journal abbreviations and non-ISI sources | Citations |
|---|---|
| *J Phys Chem-US* | 54,139 |
| *Phys Rev* | 32,352 |
| *Biochim Biophys Acta* | 26,108 |
| *Communication* | 22,062 |
| *Am J Physiol* | 14,716 |
| *Unpub* | 14,020 |
| *Am J Med Genet* | 13,467 |
| *J Bone Joint Surg* | 13,405 |
| *J Biomed Mater Res* | 12,962 |
| *J Chem Soc Perk T 1* | 11,870 |
| *Am Rev Respir Dis* | 11,033 |
| *P Soc Photo-Opt Ins* | 10,817 |
| *Acta Metall Mater* | 10,310 |
| *Mmwr-Morbid Mortal W* | 10,208 |

**Table 4**: Non-ISI sources and incorrect journal abbreviations with more than 10,000 citations in the *JCR* 2005.

In addition to the effects of the selection process on the deselected journals, the otherwise impressively rich database contains also a lot of error. Table 4 shows the fourteen "journal abbreviations" among the non-ISI sources which obtained more than 10,000 citations in 2005. With its 54,139 citations, the *J Phys Chem-US* would belong to the top-50 journals of the database if it were included. However, this journal is included in the ISI-database under the abbreviations *J Phys Chem A* and *J Phys Chem B* with 32,086 and 59,826 citations, respectively. For some journals, however, the different spellings in the references may have large implications. Bornman *et al.* (2007, at p. 105) found 21.5% overestimation of the impact factor of *Angewandte Chemie* in 2005 because of authors providing references to both the German and international editions of this journal (Marx, 2001). Increasing a journal's impact factor has become an industry in itself (Li, 2006; Park & Leydesdorff, forthcoming).

While it may nevertheless be feasible (in some cases more than in others) to delineate journal sets which provide a fair representation of a specialty, institutional units of analysis are almost never confined to a single and relatively homogenous journal set. One might even put a question mark in terms of a unit's longer-term perspective if it were publishing exclusively in a highly specialized set. Policy-makers may wish to encourage researchers to take more risks by funding "interdisciplinary" programs, but an average researcher has only a single chance of making a career in science. Trade-offs between funding opportunities and intellectual perspectives remain difficult for individual scholars, research groups, and societal stakeholders.

Where does this leave us with respect to political and managerial incentives for evaluations? Scientometric evaluations have obtained a function in public policy and the management of R&D. Policy makers and management may be inclined to opt for choices based on relevant information—and probably one should. Indicators can function to inform the various discourses. However, one should be very careful not to throw out the baby with the bathwater on the basis of normative assumptions like the expected scale-effects of large-scale concentrations of R&D on productivity (Adams & Smith, 2003; Von Tunzelman *et al.*, 2003). More often than not, one is able to generate evidence which points in another direction. The emerging knowledge-based economy may have more need to stimulate variation than to increase selection pressures (Ashby, 1958; Bruckner *et al*., 1994; Leydesdorff, 2006b).

**Acknowledgement**

I am grateful to Bill Page, the Editor of *Science & Public Policy,* for providing me with relevant data. I am also grateful to three anonymous referees.